\documentclass[aps,preprint
,showpacs
,groupedaddress
]
{revtex4-1}
\usepackage{amsmath,amsthm,amssymb}
\usepackage{array}
\usepackage{graphicx}
\usepackage{subfig}
\usepackage{fancyhdr}
\usepackage{color}
\usepackage{extarrows}
\usepackage{enumerate}
\usepackage{cleveref}

\begin{document}

\title{Zero Mass Limit and Its Experimental Test\\
\small{Comment on ``Brownian colloidal particles: Ito, Stratonovich, or a different stochastic interpretation"}}

\author{Ruoshi Yuan}
\affiliation{Department of Computer Science and Engineering Shanghai Jiao Tong University, Shanghai, 200240, China}
\author{Ping Ao}
\email{aoping@sjtu.edu.cn}
\affiliation{Shanghai Center for Systems Biomedicine and Department of Physics\\Shanghai Jiao Tong University, Shanghai, 200240, China}

\date{\today}

\begin{abstract}
J.M. Sancho [Phys. Rev. E \textbf{84}, 062102 (2011)] analyzed two stochastic interpretations on a recent experiment [Phys. Rev. Lett. \textbf{104}, 170602 (2010)] of Brownian colloidal particles. The author asserted that the stochastic interpretation ``obtained by simply setting the acceleration equal to zero" should not be taken and that the zero-mass limit interpretation of the experimental data would not be physically correct.
In this Comment we show that Sancho's analysis is incomplete in that it pre-excludes zero mass limit and hence his assertions are incorrect.  Our reasoning will be both mathematical and physical.

\end{abstract}
\maketitle

The starting point of Sancho's paper \cite{PhysRevE.84.062102} is the classical Newtonian equation ${\bf F} = m {\bf a}$:
\begin{align}
\label{newton}
   - \lambda(z)\dot{z}-V'(z)+g(z)\xi(t) = m\ddot{z}~,
\end{align}
where the mass $m$ and the friction coefficient $\lambda$ are two physical parameters which intrinsically exist. First, we should point out that the setting of $m$ as unity in Sancho's paper \cite{PhysRevE.84.062102} generally pre-excludes the discussion of zero mass limit: Mathematically, the zero mass limit $m\to 0$ is a different limit process to the overdamped (Smoluchowski) limit $\lambda/m\to\infty$. Fortunately for Sancho, his setting of mass to be unity is not essential in 1-d, which we will return in the end.

Mathematically it is already clear that Sancho's setting mass to be unity is generally inappropriate. The further important question is that whether or not the zero-mass limit is mathematically meaningful.
Indeed, by taking zero mass limit on Eq.~(1), we obtain
\begin{align}
\label{equilibrium}
   -\lambda(z)\dot{z}-V'(z)+g(z)\xi(t) = 0~,
\end{align}
which is the same as the Eq.~(5) in \cite{PhysRevE.84.062102} by a common factor $\lambda(z)$. Such limiting is deceptively simple, though it is a singular perturbation situation. Now, with precise expressions the important mathematical question becomes that, is there a mathematical procedure for Eq.~(2) which would lead to the same results as from Eq.~(1). Here we should admit that Sancho is correct in stating that the usual Ito or Stratonovich procedure on Eq.~(2) would not. Nevertheless, such of his analysis is incomplete.
Systematic analysis has shown that there are procedures beyond those of Ito and Stratonovich \cite{ao2004potential,ao2007existence}.
Exactly, for 1-d, the stochastic interpretation of $\alpha=1$ is the appropriate choice \cite{ao2007existence};
the general $n$-dimensional cases without detailed balance have been discussed as well \cite{ao2004potential,ao2007existence}.
Hence, contrast to his assertion, Eq.~(2) (Sancho's Eq.~(5)) can be mathematically meaningful.

Having reasoned mathematically, we turn to the physical meaning of Eq.~(2), which is essential, because, after all we are dealing with physical systems. First, theoretically by setting mass to be zero, Eq.~(2) is simply the famous force balance equation ${\bf F} = 0 $. It is clear that such zero mass procedure does not require the acceleration ${\bf a} =0 $. Rather, the acceleration can be very large in this zero mass limit. It is a mystery on the origin of Sancho's assertion that Eq.~(2) is ``obtained by simply setting the acceleration equal to zero". Such misunderstanding of what done in literature may add to Sancho's incorrect reasoning.

Another important physical reasoning is whether or not the experiment \cite{volpe2010influence} realized the zero mass limit. For the colloidal particle studied in \cite{volpe2010influence},
$\rho=1.51\times 10^3~kg/m^3$, $2R=1.31\times10^{-9}~m$, then the mass of the particle is
$m= 1/6 \pi (2R)^3\rho= 1.78\times10^{-15}~kg$.
The symbol $dt$ denotes the sampling time interval in \cite{volpe2010influence}, within $dt\leq 10~ms$, the authors discover that ``the force acting on the particle can be treated as locally constant". In different experiments, they choose $dt$ around the magnitude of $1~ms$.
We use an approximated formula (from Fig.~1~(b) in \cite{volpe2010influence}) for the friction coefficient with $\eta=8.5~mPa~s$ and $z_0=700~nm$:
$\lambda(z) = 6 \pi \eta R (z+z_0)/z$.
For $z\to\infty$, we have $\lambda_\infty= 6\pi\eta R= 1.1\times10^{-7} N s/m$, then the time needed for the particle to return to equilibrium after acted by
an external force $F(z)$ is
 $t(z)= [F(z) / \lambda(z)] / [F(z) / m] = m / \lambda(z)$, $t_\infty=m/ \lambda_\infty
=  1.6\times10^{-8}~s << 1\times10^{-3}~s=1~ms\approx dt$.
Since $\lambda(z)>\lambda_\infty$, such that $t(z)<t_\infty<< dt$. Thus, during the sampling time interval the particle has been in equilibrium and the mass can be considered as zero in the experiment.

Interestingly, in the 1-d one can show both theoretically, as well as numerically from the present experiment, that the zero mass limit and overdamped limit are equivalent to each other. We should point out that in higher dimensions those two limits are not equivalent due to the absence of detailed balance. In conclusion, Sancho's analysis is incomplete and his assertion of no-zero mass limit is incorrect.

\begin{acknowledgements}
This work was supported in part by the National 973 Project No.~2010CB529200 and by the Natural Science Foundation of China No.~NFSC91029738 and No.~NFSC61073087.
\end{acknowledgements}

\section*{Appendix}
\subsection{Sancho's Counter-argument}
In my opinion a Comment should be focused on the results of the paper
and here this is not the case. Most of the comment in fact has little
connection with the paper. My observations to support this assertion
follow:

1. The main purpose of the commented paper was to clarify a recent
scientific controversy surrounding a stochastic interpretation of
recent experimental data for Brownian colloidal particles (Volpe)
which was challenged by numerical simulations (Mannella). The
conclusion of the commented paper, supported by careful analytical
calculations, was that both claims were in a sense acceptable because
there are plenty of possible stochastic interpretations given the same
Boltzmann equilibrium distribution.

2. The second point of the Sancho paper is that the claim of Mannella
can be proved following a detailed analytical procedure introduced two
decades ago, while the Volpe interpretation requires a
phenomenological assumption not based on such an analysis.

3. The abstract of the comment quotes a sequence of words from the
commented paper taken out of context and makes a non correct assertion
that would imply that Sancho simply sets the acceleration equal to
zero. In fact, the sentence that begins just before Eq. (5) in the
commented paper says exactly the opposite, and I quote: ¡°In this
regard, it must be stressed that the overdamped limit is not obtained
by simply setting the acceleration equal to zero, Eq. (5), which is
precisely the starting point in the analysis of Ref. [1].¡± (Ref. [1]
is Volpe.)

4. Summarizing, in no place in the commented paper is it said that
setting the acceleration equal to zero is mathematically incorrect.
But it is clearly stated that if this arbitrary assumption is made,
then it has to be complemented with an additional assumption (not a
conventional one) about the stochastic interpretation to arrive at the
correct equilibrium Boltzmann distribution.

5. The discussion in the comment about the physical meaning of the
limit of zero mass has nothing to do with the so called overdamped
limit, which is a well established approach for objects of finite mass
moving in a fluid under a very low Reynolds number, and which is the
entire subject of the commented paper.

6. Finally the use of adjectives such as ¡°incomplete¡±,
¡°incorrect¡±, etc., is not helpful and they are inappropriate.

Moreover I agree that a scientific discussion of the existence of a
new stochastic interpretation based in experimental data is quite
interesting.

\subsection{Further Argument}

As pointed in this Comment, Sancho's analysis pre-excludes the zero mass limit,
which leads to an incomplete discussion of limiting processes. Sancho overlooked
the systematic and careful analysis of zero mass limit in the literature. It is
worth noting that Sancho did not refute this point in his/her response. His/her
criticism was minor in our opinion and focused more on semantic issues.

Without a complete analysis, Sancho's conclusion in his/her PRE paper is not
correct. In fact, his/her Eq.(5) (and Eq.(2) in our Comment) is the only
physical choice to understand the recent experiment, and it is both physically
and mathematically natural. In our view communications here indicate that
stochastic processes are far richer than what have been previously understood.
More surprising results may still wait to be found. Open and critical attitude
is indeed needed.

\end{document}